\documentclass{moriond}

\def\be{\begin{equation}}
\def\ee{\end{equation}}
\def\bea{\begin{eqnarray}}
\def\eea{\end{eqnarray}}

\newcommand{\MeV}{\,\mathrm{MeV}}

\newcommand{\TeV}{\,\mathrm{TeV}}
\usepackage{amsmath,amssymb,mathtools}

\newcommand{\Photo}{}

\begin{document}
\vspace*{4cm}
\title{$K\to\pi\nu\bar\nu$ spectra and NA62 interpretation}

\author{Martin~Gorbahn${}^{a}$, Ulserik~Moldanazarova${}^{a,b}$, Kai~Henryk~Sieja${}^{c}$, Emmanuel~Stamou${}^{c}$, \underline{Mustafa~Tabet}${}^{c}$}

\address{${}^{a}$Department of Mathematical Sciences, University of Liverpool, Liverpool L69 3BX, UK\\
${}^{b}$Faculty of Physics and Technology, Karaganda Buketov University, 100028 Karaganda, Kazakhstan\\
${}^{c}$Fakult\"at f\"ur Physik, TU Dortmund, D-44221 Dortmund, Germany
}

\maketitle\abstracts{
Using the measured and projected invisible mass spectrum of the
$K^+\to\pi^+\nu\bar\nu$ mode, we determine the current and future constraints
within the model-independent framework of the weak effective theory at
dimension-six. We work in two different operator bases depending whether
neutrinos are Majorana or Dirac fermions. This makes it possible to
transparently incorporate mass effects of additional sterile neutrinos for all
operators.
}

\section{Introduction}
The branching ratio of $K^+\to\pi^+\nu\bar\nu$ is currently being measured at
$\mathcal{O}(35\%)$ accuracy at the NA62
experiment~\cite{NA62:2020fhy,NA62:2021zjw} while an accuracy of
$\mathcal{O}(15\%)$ is possible to be reached at the end of the runtime. With
planned and future experiments like HIKE on the table an accuracy as low as
$\mathcal{O}(5\%)$ is expected. In order to quantify the New Physics (NP)
sensitivity, this has to be compared with theoretical predictions. In the
Standard Model (SM) the decay $K^+\to\pi^+\nu\bar\nu$ is highly suppressed as it
is generated at the loop-level. The total theoretical uncertainty on the SM
prediction is $\approx 6\%$ and is dominated by the CKM input parameters, and
thus, the rare decay $K^+\to\pi^+\nu\bar\nu$ has a high sensitivity to NP.

In anticipation of this new data, we consider NP effects to all relevant weak
effective theory operators up to dimension-six for both Majorana and Dirac
neutrinos including the effect of additional massive sterile neutrinos on the
missing-mass spectrum of $K^+\to\pi^+\nu\bar\nu$. We work in two distinct
operator bases depending whether neutrinos are Majorana or Dirac fermions. This
allows for a transparent interpretation of possible NP signals in terms of
lepton-number violating or lepton-number conserving NP. In order to determine
the constraints on the relevant operators, we perform a statistical analysis in
a fully frequentist manner and incorporate the full experimental information on
the missing-mass distribution.

\section{Model Setup}
In this section we present the operators bases for the two limiting cases
of Majorana and Dirac neutrinos, respectively.

\subsection{Majorana-$\nu$ EFT}
The effective Lagrangian for the $d_i\to d_j\nu\nu$ transition and Majorana
neutrinos at dimension-six reads
\begin{align}\label{eq:lag-wet-majorana}
\mathcal{L}^{(6)}_{d_i \to d_j\nu\nu}\Big\vert_{\mathrm{Majorana}}
  = \sum_{I = \left\{ V,A \right\} ,
    \tau = \left\{ L,R \right\}}
    \sum_{f} C^{I, \tau}_f O^{I, \tau}_f
    + \left(
      \sum_{ I = \left\{ S,P,T \right\} }
      \sum_{f} C^{I,L}_f O^{I,L}_f + \mathrm{h.c.}
    \right) \,,
\end{align}
with the interaction type $I = \{V, A, S, P, T\}$ for vector, axial-vector,
scalar, pseudoscalar, and tensor, respectively, the chirality $\tau = {L, R}$ of
the fermions with flavour $f$, and the independent operators
\begin{align}
  O^{V, \, L}_{ab ij} &=\frac{1}{2}(\overline{\nu}_{Ma}\gamma_\mu\,  \nu_{Mb})(\overline{d}_i\,\gamma^\mu \,P_L\, {d}_j )\,,&
  O^{V, \, R}_{ab ij} &=\frac{1}{2}(\overline{\nu}_{Ma}\gamma_\mu\,  \nu_{Mb})(\overline{d}_i\,\gamma^\mu \,P_R\, {d}_j )\,,\\
  O^{A, \, L}_{ab ij} &=\frac{1}{2}(\overline{\nu}_{Ma}\gamma_\mu\gamma_5\, \, \nu_{Mb})(\overline{d}_i\,\gamma^\mu \,P_L\, {d}_j ) \,,&
  O^{A, \, R}_{ab ij} &=\frac{1}{2}(\overline{\nu}_{Ma}\gamma_\mu\,\gamma_5\, \nu_{Mb})(\overline{d}_i\,\gamma^\mu \,P_R\, {d}_j )\,,\\
  O^{S, \, L}_{ab ij} &=\frac{1}{2}(\overline{\nu}_{Ma} \, \nu_{Mb})(\overline{d}_i \,P_L\, {d}_j )\,,&
  O^{P, \, L}_{ab ij} &=\frac{1}{2}(\overline{\nu}_{Ma} \,i\gamma_5\, \nu_{Mb})(\overline{d}_i \,P_L\, {d}_j )\,,\\
  O^{T, \, L}_{ab ij} &=\frac{1}{2}(\overline{\nu}_{Ma} \,\sigma_{\mu\nu}\,\nu_{Mb})(\overline{d}_i \,\sigma^{\mu\nu}\,P_L\, {d}_j )\,.
\end{align}
The corresponding Wilson coefficients $C_f^{V/A,\, L/R}$ of the operators
$O^{V/A,\, L/R}$ further fulfil the condition
\begin{align}
  C_{abij}^{V/A,\, L/R} = \left( C_{baji}^{V/A,\, L/R} \right)^* \,,
\end{align}
to ensure the hermiticity of the Lagrangian in Eq.~\eqref{eq:lag-wet-majorana}.
Note that further
\begin{align}
  C_{abij}^{I,\,\tau} = \eta C_{baij}^{I,\,\tau}
  \quad\quad \text{where} \quad
  \eta = \begin{cases}
    +1, & \text{for} \quad I = A,\,S,\,P\\
    -1, & \text{for} \quad I = V,\, T \,,
  \end{cases}
\end{align}
due to the Majorana nature of the neutrinos. Altogether this amounts to a total
number of 48 independent Wilson coefficients contributing to the $s\to d\nu\nu$
transition in the case of three Majorana neutrinos.

\subsection{Dirac-$\nu$ EFT}
The effective Lagrangian for the $d_i\to d_j\nu\bar\nu$ transition and Dirac
neutrinos at dimension-six reads
\begin{align}\label{eq:lag-wet-dirac}
\mathcal{L}^{(6)}_{d_i\to d_j\nu\nu}\Big\vert_{\mathrm{Dirac}}
  = &\sum_{\tau,\tau'=\left\{L,R\right\}}
     \sum_{f} C^{V,\,\tau\tau'}_f O^{V,\,\tau\tau'}_f \notag \\
    & + \sum_{f} \left(
    C^{S,\,LL}_f O^{S,\,LL}_f +
    C^{S,\,LR}_f O^{S,\,LR}_f +
    C^{T,\,LL}_f O^{T,\,LL}_f + \mathrm{h.c.}
  \right) \,,
\end{align}
with the chiralities $\tau,\tau' = \{L, R\}$ of the neutrino and quark flavours
denoted collectively by $f$. The independent operators in the Lagrangian in
Eq.~\eqref{eq:lag-wet-dirac} read
\begin{align}
  O^{V, \,LL}_{\,ab ij} &=(\overline{\nu}_{Da}\gamma_\mu\, P_L\, \nu_{Db})(\overline{d}_i\,\gamma^\mu \,P_L\, {d}_j )\,,&
  O^{V, \,LR}_{\,ab ij} &=(\overline{\nu}_{Da}\gamma_\mu\, P_L\, \nu_{Db})(\overline{d}_i\,\gamma^\mu \,P_R\, {d}_j )\,,\\
  O^{V, \,RL}_{\,ab ij} &=(\overline{\nu}_{Da}\gamma_\mu\, P_R\, \nu_{Db})(\overline{d}_i\,\gamma^\mu \,P_L\, {d}_j )\,,&
  O^{V, \,RR}_{\,ab ij} &=(\overline{\nu}_{Da}\gamma_\mu\, P_R\, \nu_{Db})(\overline{d}_i\,\gamma^\mu \,P_R\, {d}_j )\,,\\
  O^{S,\,LL}_{\,ab ij} &=(\overline{\nu}_{Da} \, P_L\, \nu_{Db})(\overline{d}_i \,P_L\, {d}_j )\,,&
  O^{S,\,LR}_{\,ab ij} &= (\overline{\nu}_{Da} \, P_L\, \nu_{Db})(\overline{d}_i \,P_R\, {d}_j )\,,\\
  O^{T,\,LL}_{\,ab ij} &= (\overline{\nu}_{Da} \,\sigma_{\mu\nu}\,P_L\,\nu_{Db})(\overline{d}_i \,\sigma^{\mu\nu}\,P_L\, {d}_j )\,.&&
\end{align}
Here the hermiticity condition on the corresponding Wilson coefficients read
\begin{align}
  O_{abij}^{V,\,\tau\tau'} = \left( O_{baji}^{V,\,\tau\tau'} \right)^* \,.
\end{align}
Altogether this amounts to a total number of 90 independent Wilson coefficients
contributing to the $s\to d\nu\bar\nu$ transition in the case of three Majorana
neutrinos.

\section{New Physics Sensitivities and Correlations}
We first show the differential distributions for the Dirac and Majorana
effective theory with three light neutrinos $m_{\nu,1}\approx m_{\nu,2}\approx
m_{\nu,3}\approx 0$, and the case of one additional sterile neutrino with mass
$m_{\nu,4} = 50\MeV$ within the Majorana effective theory with three massless
neutrinos. Further, we present the correlations between new diagonal Dirac
neutrino couplings and the current and projected limits on the Wilson
coefficients for the case of an additional sterile neutrino. Further scenarios
can be found in Ref.~\cite{Gorbahn:2023juq}. The limits always correspond to the
NP interactions, i.e. in addition to the SM contribution. In
case of possible interference terms with the SM, the NP phase is taken to be
aligned to the SM phase.

\subsection*{Single-operator fits}
\begin{figure}
  \centering
  \includegraphics[width=0.95\textwidth]{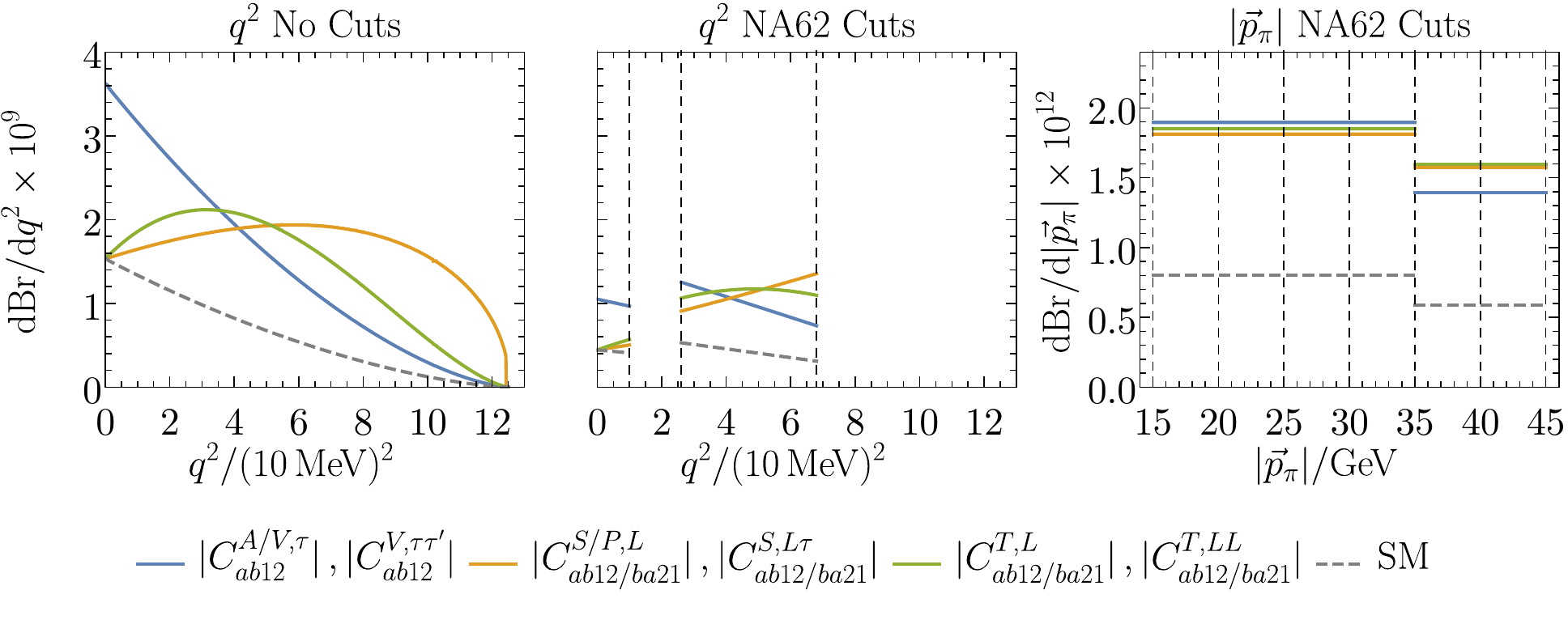}
  \caption{
    Differential distributions of $\mathrm{Br}(K\to\pi\nu\nu)$ for different NP
    scenarios containing three massless Majorana or Dirac neutrinos, see text
    for details.
  }
  \label{fig:dists}
\end{figure}

\begin{figure}
  \centering
  \includegraphics[width=0.95\textwidth]{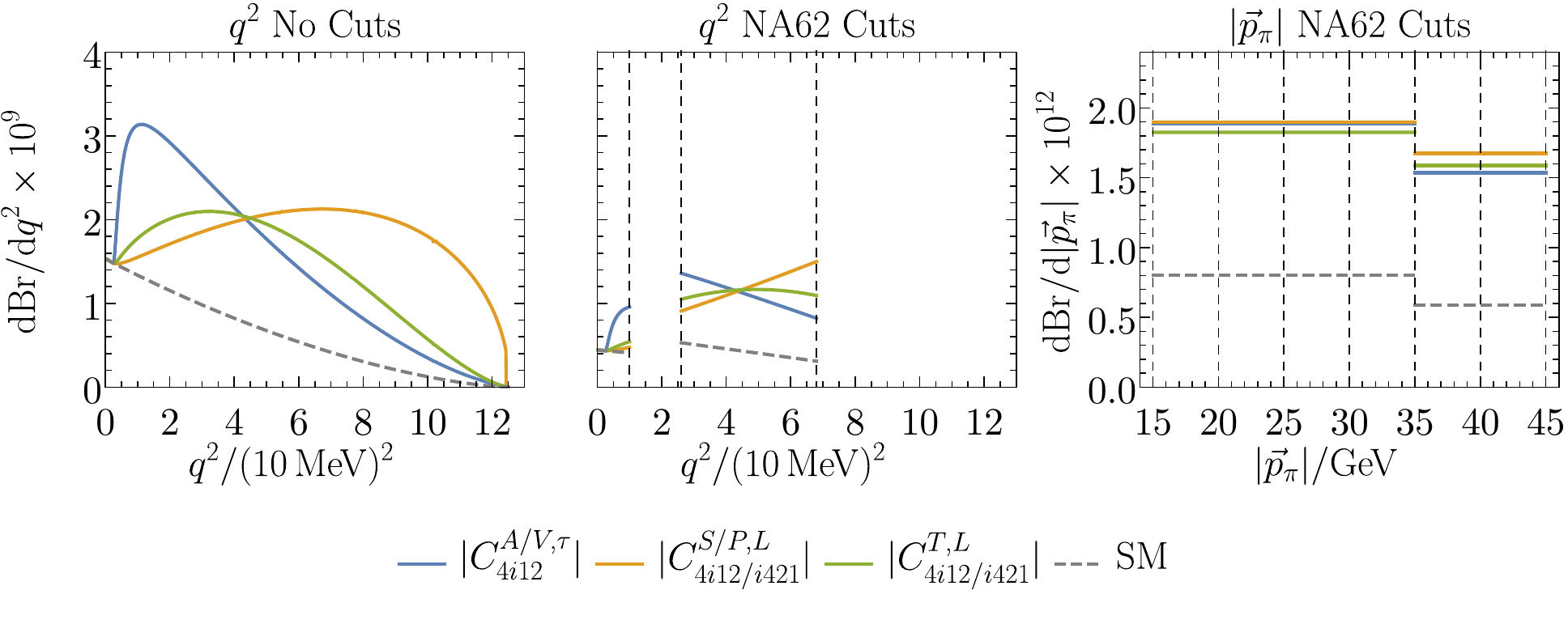}
  \caption{
    Differential distributions of $\mathrm{Br}(K\to\pi\nu\nu)$ for different NP
    scenarios containing one massive sterile neutrino with mass,
    $m_{\nu,4}=50\MeV$, see text for details.
  }
  \label{fig:dists2}
\end{figure}

Here we only present the limits for the case of a Majorana effective theory with
three light neutrinos and an additional sterile neutrino. The Dirac and Majorana
scenarios with three massless neutrinos can be found in Ref.~\cite{Gorbahn:2023juq}.
Nevertheless, we first show the differential distributions for the Dirac and
Majorana case in Fig.~\ref{fig:dists} and the case of an additional sterile
neutrino with mass $m_{\nu,4} = 50\MeV$ in Fig.~\ref{fig:dists2} for all independent
Wilson coefficients.
In both figures the value of the Wilson coefficients are chosen such that
they saturate the current experimental limit at $90\%$ CL.
As it can be seen in Fig.~\ref{fig:dists}, the case of Majorana and Dirac neutrinos
is experimentally indistinguishable.

In the left panel of both figures, we show the differential distribution
$\mathrm{dBr}(K^+\to\pi^+\nu\nu)/\mathrm{d}q^2$ while in the middle panels, we
show the same but integrated over the pion-momentum signal region of NA62. The
right panels show the distributions differential in the pion momentum
$|\vec{p}_\pi|$ in the NA62 lab frame integrated over the $q^2$ signal region of
NA62. As it can be seen in both figures, a binning in the missing-momentum squared $q^2$
is able to distinguish between the different interaction structures but not between
the Majorana or Dirac nature of the neutrinos compared to a binning in the pion momentum
$|\vec{p}_\pi|$ as currently employed by the NA62 experiment.

\begin{figure}
  \centering
  \includegraphics[width=0.95\textwidth]{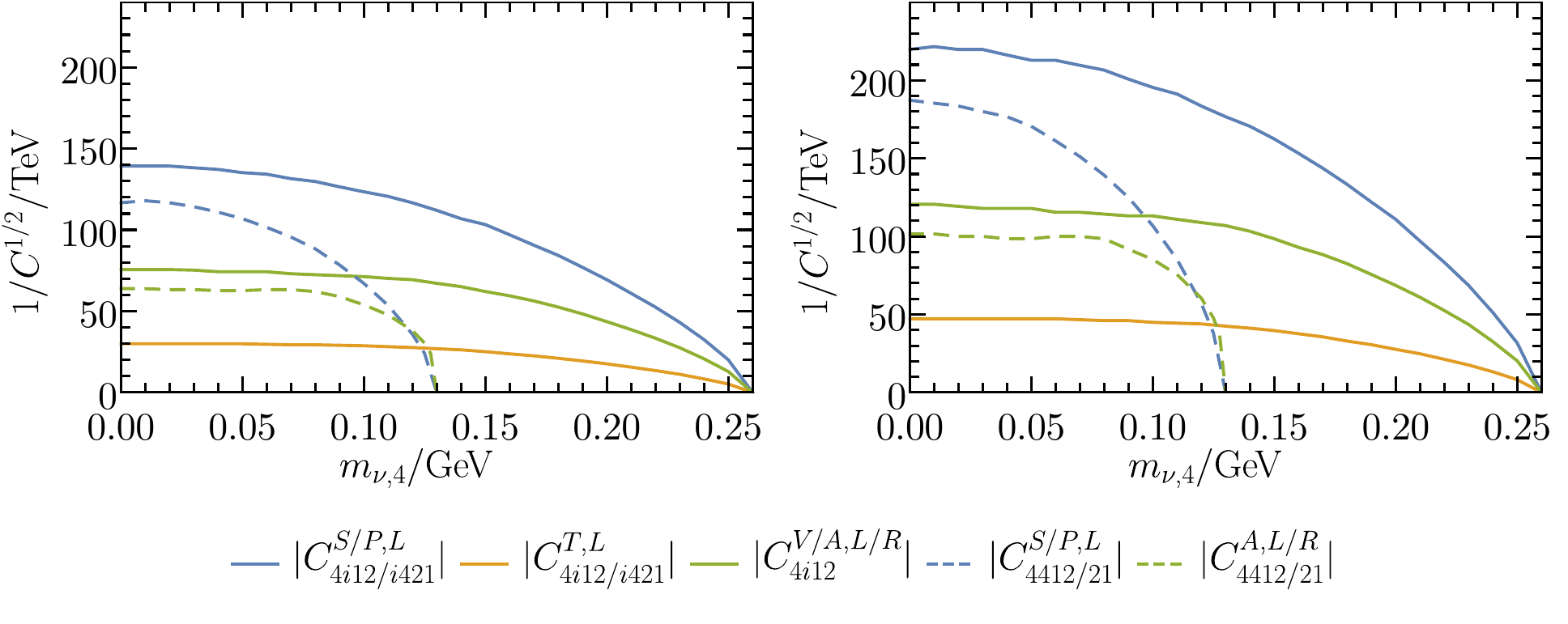}
  \caption{
    Current (left) and future (right) lower limits on $1/\sqrt{C}$ at 90\% CL
    for the scenarios with an additional massive sterile neutrino interacting
    via one of the given NP Wilson coefficients as a function of its mass.
  }
  \label{fig:constraints-sterile-maj}
\end{figure}

Finally, we show the resulting limits on the Wilson coefficients as a function of the
mass of an additional sterile neutrino in Figure~\ref{fig:constraints-sterile-maj}.
In the left panel of Figure~\ref{fig:constraints-sterile-maj}, we show the current
constraints while in the right panel we show the projected ones for the HIKE experiment.

\subsection*{Multi-operator correlations}
\begin{figure}
  \centering
  \includegraphics[width=0.95\textwidth]{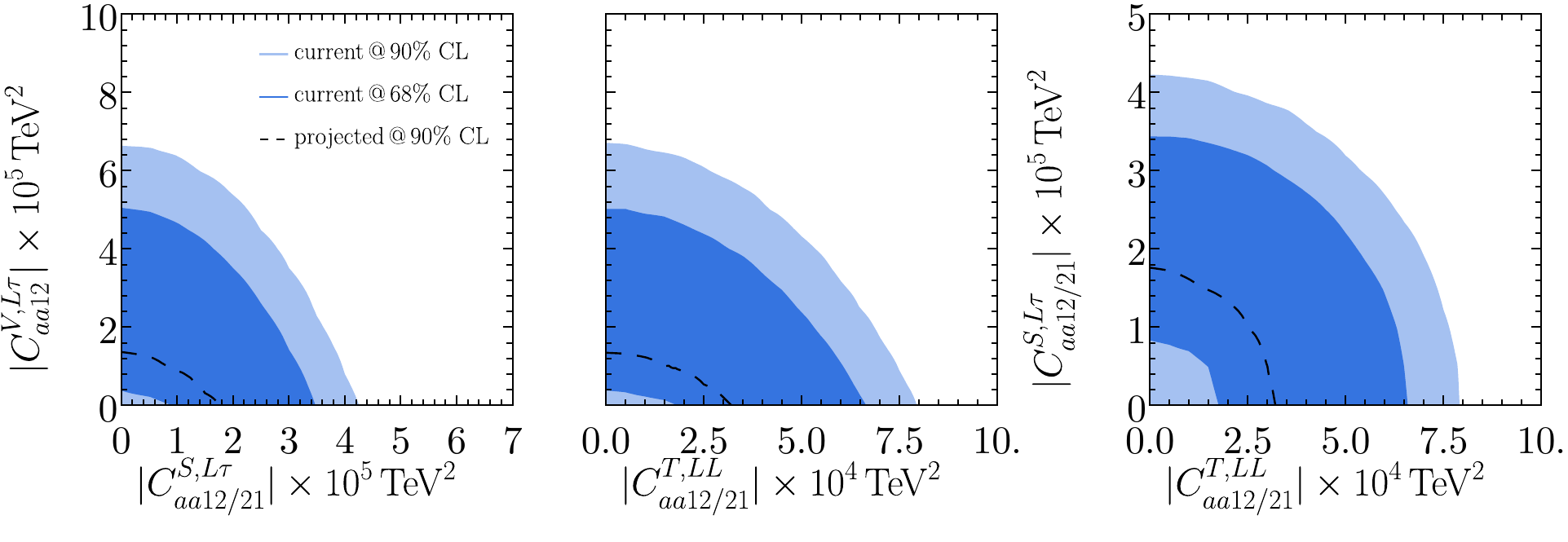}
  \caption{
    Allowed parameter space at 68\% CL (dark blue) and 90\% CL (light blue) in
    the plane of two NP Wilson coefficients, see text for details.
  }
  \label{fig:correlations}
\end{figure}

Here we study the correlation between different NP operators
in particular the case of new diagonal Dirac interactions on top of the
already present SM contribution. All NP phases are aligned to the SM phase
in order to avoid new CP-violating contributions. Additionally, all new interactions
are taken to be flavour universal and are switched on at the same time, i.e.
$C^{V,\,LL}_{1112} = C^{V,\,LL}_{2212} = C^{V,\,LL}_{3312}$.
The results are shown in Fig.~\ref{fig:correlations} where the dark and light blue regions
correspond to the allowed regions at $68\%$ and $90\%$ CL. The dashed black line
corresponds to the projected expected limit at $90\%$ CL at HIKE assuming a SM like
measurement.

\section{Conclusion}
The rare decay mode $K\to\pi\nu\nu$ currently probes NP scales up to
$\mathcal{O}(100\TeV)$ and with upcoming and planned experiments the constraints
are expected to reach scales as high as $\mathcal{O}(300\TeV)$. Furthermore, all
independent operators contributing to the $q_i\to q_j \nu\nu$ ($i\neq j$)
transition have been classified. While here we only have shown the constraints
exemplary for the case of new diagonal flavour universal Dirac interactions, the
sensitivities of all independent operators and correlations between them can be
found in Ref.~\cite{Gorbahn:2023juq}. There, the expected constraints on the NP
scale range from $\mathcal{O}(40\TeV)$ for tensor operators to
$\mathcal{O}(300\TeV)$ for axial-vector Majorana and the here analysed vector
Dirac operators. We further analysed the effect of an additional massive sterile
neutrino on the distributions and determined the current and future experimental
sensitivity on constraining the corresponding Wilson coefficients as a function
of the sterile neutrino mass.

\section*{Acknowledgments}
This contribution to the EW session of the 58th Rencontres de Moriond 2024 is
based on Ref.~\cite{Gorbahn:2023juq}.

\end{document}